\documentclass[sigconf,screen]{acmart}

\usepackage{import}
\usepackage[normalem]{ulem}
\usepackage{natbib}
\usepackage{blindtext}  %
\usepackage{enumitem}   %
\usepackage{hyperref}   %
\usepackage{cleveref}   %

\makeatletter
\let\MYcaption\@makecaption
\makeatother

\usepackage[font=footnotesize]{subcaption}

\makeatletter
\let\@makecaption\MYcaption
\makeatother

\usepackage[multiple]{footmisc} %

\usepackage{graphicx}
\graphicspath{{images/}{../images/}}
\usepackage{listings}
\usepackage{textcomp}
\usepackage{color}
\definecolor{maroon}{rgb}{0.5,0,0}
\definecolor{darkgreen}{rgb}{0,0.5,0}

\newcommand{\beginlstdelim}[3]
{%
  \def\endlstdelim{#2\egroup}%
  \ttfamily#1\bgroup\color{#3}\aftergroup\endlstdelim%
}

\lstdefinestyle{xml_style}
{
  morestring=[b]",
  moredelim=*[s][\bfseries\color{maroon}]{<}{>},
  moredelim=*[s][\bfseries\color{maroon}]{</}{>},
  morecomment=[s]{<?}{?>},
  morecomment=[s]{<!--}{-->},
}

\lstdefinestyle{json_style}
{
  morestring=[b]",
  comment=[l]{//},
  keywordstyle=\bfseries\color{maroon}
}

\lstdefinelanguage{JSON}
{
  style=json_style
}

\lstdefinelanguage{XML}
{
  style=xml_style
}
\usepackage{booktabs}
\usepackage{multirow}
\usepackage{makecell}

\usepackage{xcolor}
\newcommand{\todo}[2][]{#1} %
  \usepackage{balance}

\usepackage{tikz}
\newcommand*\circled[1]{\tikz[baseline=(char.base)]{
            \node[shape=circle,draw,inner sep=1pt] (char) {#1};}}
\begin{document}

\def\demolink{https://bit.ly/2YKeYhE}
\def\demogooglelink{http://bit.ly/a2i2-threshy}

\title{Threshy: Supporting Safe Usage of Intelligent Web Services}

\settopmatter{authorsperrow=4}

\author{Alex Cummaudo}
\email{ca@deakin.edu.au}
\orcid{0000-0001-7878-6283}
\affiliation{%
\department{Applied Artificial Intel. Inst.}
\institution{Deakin University}
\city{Geelong}
\country{Australia}
}

\author{Scott Barnett}
\email{scott.barnett@deakin.edu.au}
\orcid{0000-0002-3187-4937}
\affiliation{%
\department{Applied Artificial Intel. Inst.}
\institution{Deakin University}
\city{Geelong}
\country{Australia}
}

\author{Rajesh Vasa}
\email{rajesh.vasa@deakin.edu.au}
\orcid{0000-0003-4805-1467}
\affiliation{%
\department{Applied Artificial Intel. Inst.}
\institution{Deakin University}
\city{Geelong}
\country{Australia}
}

\author{John Grundy}
\email{john.grundy@monash.edu}
\orcid{0000-0003-4928-7076}
\affiliation{%
\department{Faculty of Inf. Tech.}
\institution{Monash University}
\city{Clayton}
\country{Australia}
}

\begin{abstract}
Increased popularity of `intelligent' web services provides end-users with machine-learnt functionality at little effort to developers. However, these services require a decision threshold to be set which is dependent on problem-specific data. Developers lack a systematic approach for evaluating intelligent services and existing evaluation tools are predominantly targeted at data scientists for pre-development evaluation. This paper presents a workflow and supporting tool, Threshy, to help \textit{software developers} select a decision threshold suited to their problem domain. Unlike existing tools, Threshy is designed to operate in multiple workflows including pre-development, pre-release, and support. Threshy is designed for tuning the confidence scores returned by intelligent web services and does not deal with hyper-parameter optimisation used in ML models. Additionally, it considers the financial impacts of false positives.
Threshold configuration files exported by Threshy can be integrated into client applications and monitoring infrastructure. Demo: \url{\demolink}.
\end{abstract}

\ccsdesc[500]{Information systems~Web services}
\ccsdesc[500]{Computing methodologies~Artificial intelligence}
\ccsdesc[500]{Software and its engineering~Designing software}
\ccsdesc[300]{Software and its engineering~Software post-development issues}

\keywords{intelligent services, tooling, thresholding, decision theory}

\renewcommand{\shortauthors}{Alex Cummaudo et al.}

\copyrightyear{2020}
\acmYear{2020}
\setcopyright{acmlicensed}\acmConference[ESEC/FSE '20]{Proceedings of the 28th ACM Joint European Software Engineering Conference and Symposium on the Foundations of Software Engineering}{November 8--13, 2020}{Virtual Event, USA}
\acmBooktitle{Proceedings of the 28th ACM Joint European Software Engineering Conference and Symposium on the Foundations of Software Engineering (ESEC/FSE '20), November 8--13, 2020, Virtual Event, USA}
\acmPrice{15.00}
\acmDOI{10.1145/nnnnnnn.nnnnnnn}
\acmISBN{nnn-n-nnnn-nnnn-1/20/11}
\maketitle

\section{Introduction}

Machine learning (ML) algorithm adoption is increasing in modern software. End users routinely benefit from machine-learnt functionality through personalised recommendations \citep{covington2016deep}, voice-user interfaces \citep{myers2018patterns}, and intelligent digital assistants \citep{boyd2018just}. The easy accessibility and availability of intelligent web services\footnote{Such as Azure Computer Vision (\url{https://azure.microsoft.com/en-au/services/cognitive-services/computer-vision/}), Google Cloud Vision (\url{https://cloud.google.com/vision/}), or Amazon Rekognition (\url{https://aws.amazon.com/rekognition/}).} is contributing to their adoption. These intelligent web services simplify the development of ML solutions as they (i)~do not require specialised ML expertise to build and maintain AI-based solutions, (ii)~abstract away infrastructure related issues associated with ML \citep{Sculley2015, Arpteg2018}, and (iii)~provide web APIs for ease of integration.

However, unlike traditional web services, the functionality of these \textit{intelligent services} is dependent on a set of assumptions unique to ML~\citep{Cummaudo:2019icsme}. These assumptions are based on the data used to train ML algorithms, the choice of algorithm, and the choice of data processing steps---most of which are not documented. For developers, these assumptions mean that the performance characteristics of an intelligent service in any particular application problem domain is not fully knowable. Intelligent services represent this uncertainty through a confidence value associated with their predictions.

\begin{figure}[t]
    \includegraphics[width=\linewidth]{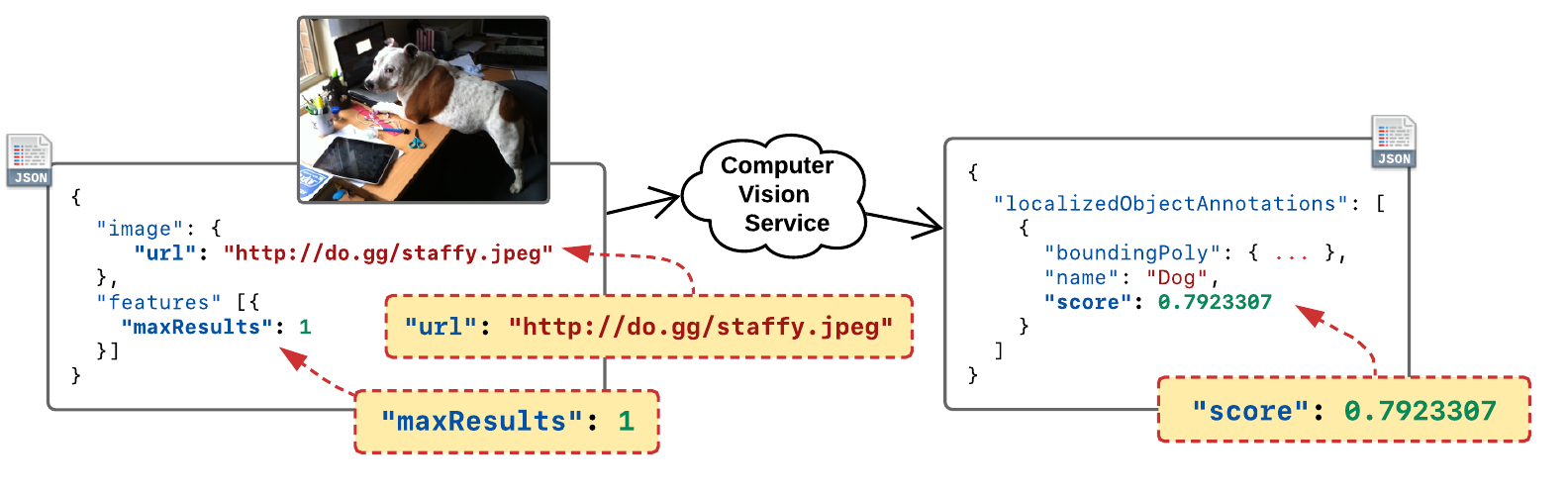}
    \caption{Request and response for an intelligent computer vision web service with only three configuration parameters: the image's \texttt{url}, \texttt{maxResults} and \texttt{score}.}
    \label{fig:dog-example}
\end{figure}

As an example, consider \cref{fig:dog-example}, which illustrates an image of a dog uploaded to a real computer vision service. Developers have very few configuration parameters in the upload payload (\texttt{url} of the image to analyse and \texttt{maxResults} the number of objects to detect). The JSON output payload returns the confidence value via a \texttt{score} field (0.792), the bounding box and a ``dog'' label. Developers can only work with these parameters; unlike hyper-parameter optimisation available to ML creators, who can configure the internal parameters of the algorithm while training a model. Given the structure of the abstractions, developers have no insight into which hyper-parameters are used or the algorithm selected and cannot tune the underlying trained model when using an intelligent service. %
Thus an evaluation procedure must be followed as a part of using an intelligent service for an application to work with and tune the output confidence values for a given input set.

\begin{figure}[t]
    \centering
    \includegraphics[width=.9\linewidth]{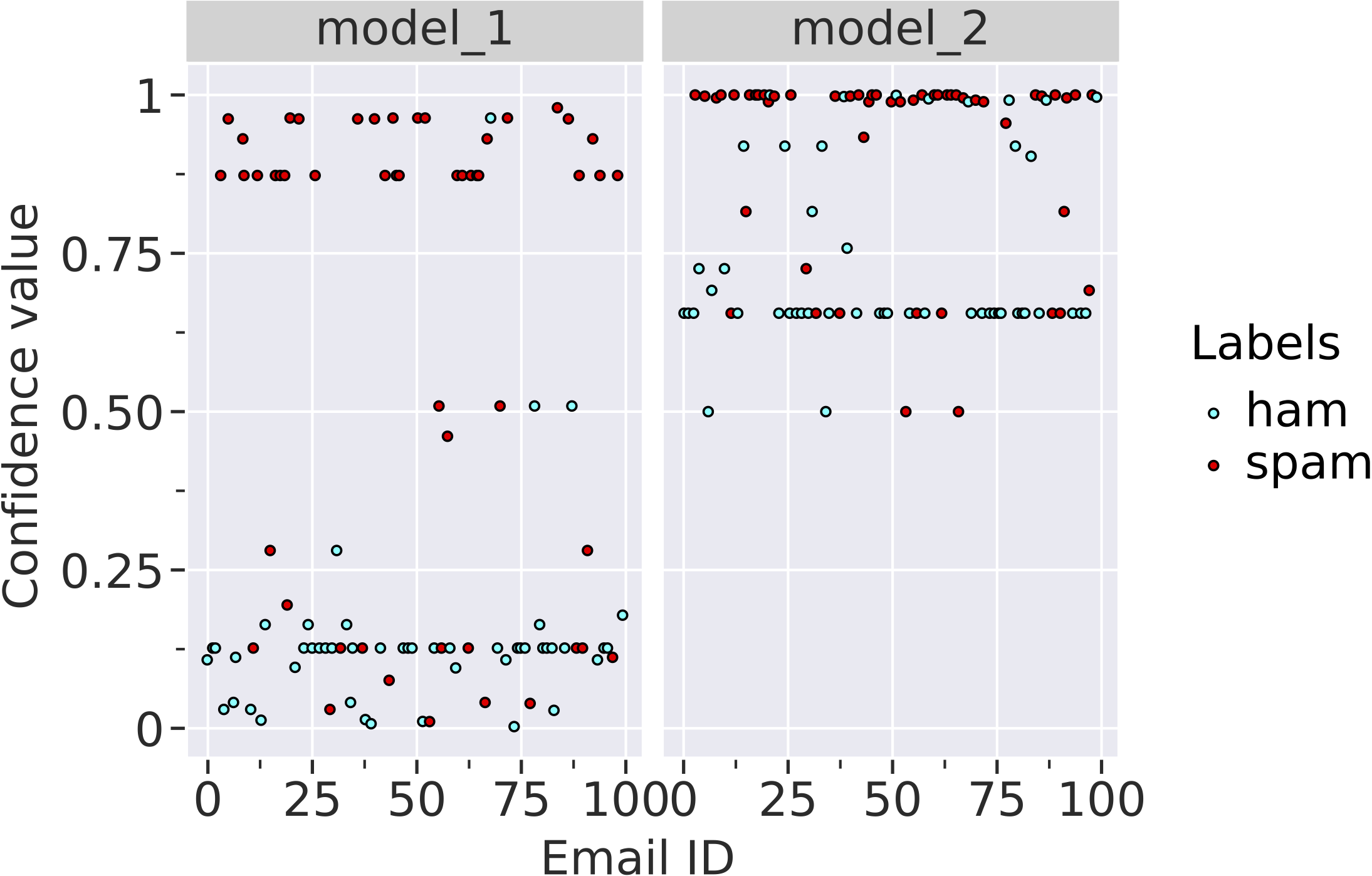}
    \caption{Predictions for 100 emails from two spam classifiers. Decision thresholds are classifier-dependent:  a single threshold for both classifiers is \textit{not} appropriate as ham emails are clustered at 0.12 (model\_1) and at 0.65 (model\_2). Developers must evaluate  performance for \textit{both} thresholds.}
    \label{fig:example}
\end{figure}

A typical evaluation process involves a test data set (curated by the developers using the intelligent service) that is used to determine an appropriate threshold. Choice of a decision threshold is a critical element of the evaluation procedure \citep{hardt2016equality}. This is especially true for classification problems such as detecting if an image contains cancer. Simple approaches to selecting a threshold are often insufficient, as highlighted in Google's ML course: \textit{``It is tempting to assume that [a] classification threshold should always be 0.5, but \textbf{thresholds are problem-dependent, and are therefore values that you must tune}.''}\footnote{See \url{https://bit.ly/36oMgWb}.}

As an example consider the predictions from two email spam classifiers shown in \Cref{fig:example}. The predicted safe emails, `ham', are in two separate clusters (a simple threshold set to approx. 0.2 for model 1 and 0.65 for model 2, indicating that different decision thresholds may be required depending on the classifier. Also note that some emails have been misclassified; how many depends on the choice of decision threshold. An appropriate threshold considers factors outside algorithmic performance, such as financial cost and impact of wrong decisions. To select an appropriate decision threshold, developers using intelligent services need approaches to reason about and consider trade-offs between competing \textit{cost factors}. These include impact, financial costs, and maintenance implications. Without considering these trade-offs, sub-optimal decision thresholds will be selected.

The standard approach for tuning thresholds in classification problems involve making trade-offs between the number of false positives and false negatives using the receiver operating characteristic (ROC) curve. However, developers (i)~need to realise that this trade-off between false positives and false negatives is a data dependent optimisation process \citep{sculley2011detecting}, (ii)~often need to develop custom scripts and follow a trial-and-error based approach to determine a threshold, (iii)~must have appropriate statistical training and expertise, and (iv)~be aware that multi-label classification require more complex optimisation methods when setting label specific costs. However, current intelligent services do not sufficiently guide or support software engineers through the evaluation process, nor do they make this need clear in the documentation.

In this paper we present \textit{\bfseries Threshy}\footnote{Threshy is available for use at \url{\demogooglelink}}, a tool to assist developers in selecting decision thresholds when using intelligent services. The motivation for developing Threshy arose from our work across a set of industry projects, \todo[and is an implemented example of the threshold tuner component presented in our complementing ESEC/FSE 2020 architecture tactic publication \citep{Cummaudo:2020fse}]{Mention it is complementary piece.}. \todo[While Threshy has been designed to specifically handle pre-trained classification ML models where the hyperparameters cannot be tuned, the overall conceptual design serves as inspiration for general model calibration.]{C16: Design of Threshy}  Unlike existing tooling (see \cref{sec:related-work}), \textbf{Threshy serves as a means to up-skill and educate software engineers in selecting machine-learnt decision thresholds}, for example, on aspects such as confusion matrices. We re-iterate that the end-users of Threshy are software engineers and not data scientists---Threshy is \uline{not} designed for hyper-parameter tuning of models, but for threshold tuning to use intelligent web services more robustly where internal models are not exposed. Threshy provides a visually interactive interface for developers to fine-tune thresholds and explore trade-offs of prediction hits/misses. This exposes the need for optimisation of thresholds, which is dependent on particular use cases.

\begin{figure}
    \centering
    \includegraphics[width=.9\linewidth]{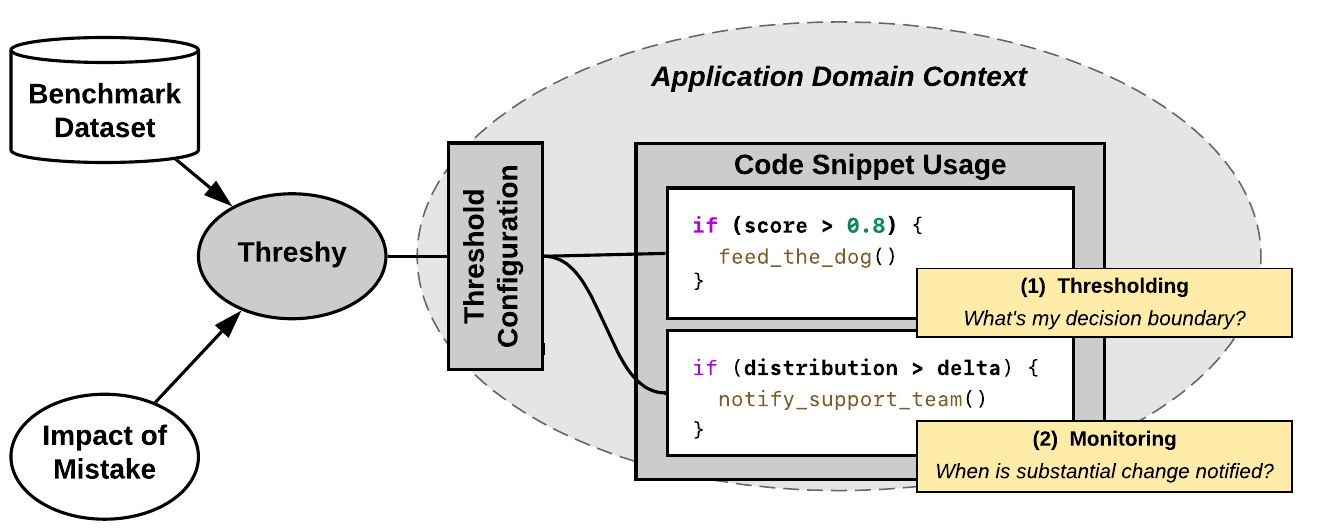}
    \caption{Threshy supports two key aspects for intelligent web services: threshold selection and monitoring.}
    \label{fig:usage-thresholding-monitoring}
\end{figure}

Threshy improves developer productivity through automation of the threshold selection process by leveraging an optimisation algorithm to propose thresholds. \Cref{fig:usage-thresholding-monitoring} illustrates the two key aspects by which Threshy supports developer's application domain context. Developers input a representative dataset of their application data (a benchmark dataset) in addition to cost factors to Threshy. Threshy's output helps developers select appropriate thresholds while considering different cost factors and can be used to monitor the evolution of an intelligent service. Developers also benefit from the workflow implemented in Threshy by providing a reproducible procedure for testing and tuning thresholds for any category of classification problem (binary, multi-class, and multi-label).
The output, is a configuration file that can be integrated into client applications ensuring that the thresholds can be updated without code changes, and continuously monitored in a production setting.

\section{Motivating Example}
\label{sec:motivating-example}

\begin{figure}[t]
  \centering
  \begin{subfigure}[b]{0.2\linewidth}
    \includegraphics[width=\textwidth]{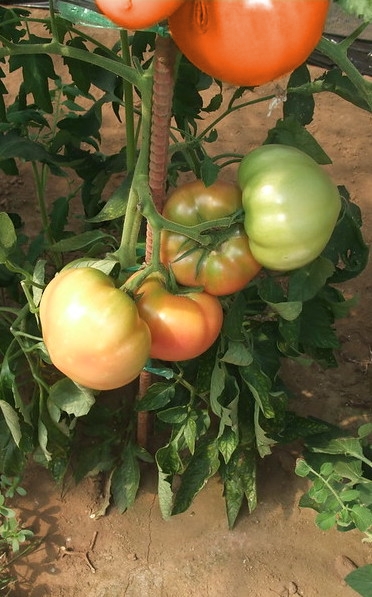}
  \end{subfigure}
  \hspace{\fill}
  ~
  \begin{subfigure}[b]{0.2\linewidth}
    \includegraphics[width=\textwidth]{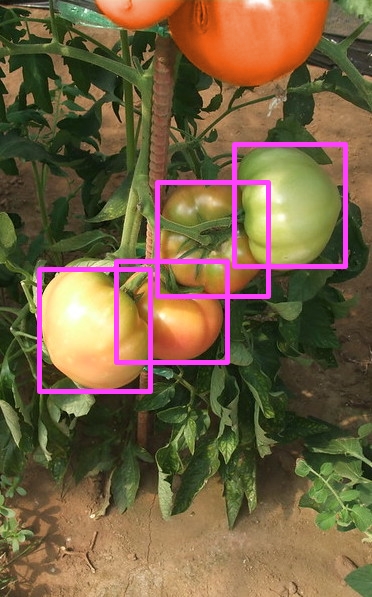}
  \end{subfigure}
  \hspace{\fill}
  ~
  \begin{subfigure}[b]{0.25\linewidth}
    \includegraphics[width=.5\textwidth]{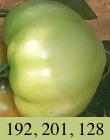}~
    \includegraphics[width=.5\textwidth]{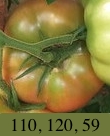}

    \includegraphics[width=.5\textwidth]{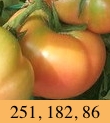}~
    \includegraphics[width=.5\textwidth]{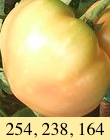}
  \end{subfigure}
  \caption{Pipeline of Nina's harvesting robot. \textit{Left:}~Photo from harvesting robot's webcam. \textit{Centre:}~Classification detecting different types of tomatoes. \textit{Right:}~Binary classification for ripeness (ripe/unripe) based on (R, G, B values).}%
  \label{fig:tomatoes}
\end{figure}

As a motivating example consider Nina, a fictitious developer, who has been employed by Lucy's Tomato Farm to automate the picking of tomatoes from their vines (when ripe) using computer vision and a harvesting robot. Lucy's Farm grow five types of tomatoes (roma, cherry, plum, green, and yellow tomatoes). Nina's robot---using an attached camera---will crawl and take a photo of each vine to assess it for harvesting.
Nina's automated harvester needs to sort picked tomatoes into a respective container, and thus several business rules need to be encoded into the prediction logic to sort each tomato detected based on its \textit{ripeness} (ripe or not ripe) and \textit{type of tomato} (as above).
Nina uses a two-stage pipeline consisting of a multi-class and a binary classification model. She has decided to evaluate the viability of cloud based intelligent services and use them if operationally effective. \Cref{fig:tomatoes} illustrates the pipeline used:
\begin{enumerate}
    \item \textbf{Classify tomato `type'.} This stage uses an object localisation service to detect all tomato-like objects in the frame and classifies each tomato into one of the following labels: \texttt{[`roma',`cherry',`plum',`green',`yellow',`unknown']}.
    \item \textbf{Assess tomato `ripeness'.} This stage uses a crop of the localised tomatoes from the original frame to assess the crop's colour properties (i.e., average colour must have R > 200 and G < 240). This produces a binary classification to deduce whether the tomato is ripe or not.
\end{enumerate}

Nina only has a minimal appreciation of the evaluation method to use for off-the-shelf computer vision (classification) services. She also needs to consider the financial costs of misclassifying either the tomato type or the ripeness. Missing a few ripe tomatoes isn't a significant concern as the robot travels the field twice a week during harvest season. However, picking an unripe tomato is expensive as Lucy cannot sell them. Therefore, Nina needs a better (automated) way to assess the performance of the service and set optimal thresholds for her picking robot, to maximise profit.

To assist in developing Nina's pipeline, Lucy sampled a section of 1000 tomatoes by taking a photo of each tomato, manually labelling its type, and assessing whether the vine was  \texttt{`ripe'} or \texttt{`not\_ripe'}. Nina ran the labelled images through an intelligent service, with each image having a predicted type (multi-class) and ripeness (binary), with respective confidence values.

\begin{figure*}[t!]
    \centering
    \includegraphics[width=.7\linewidth]{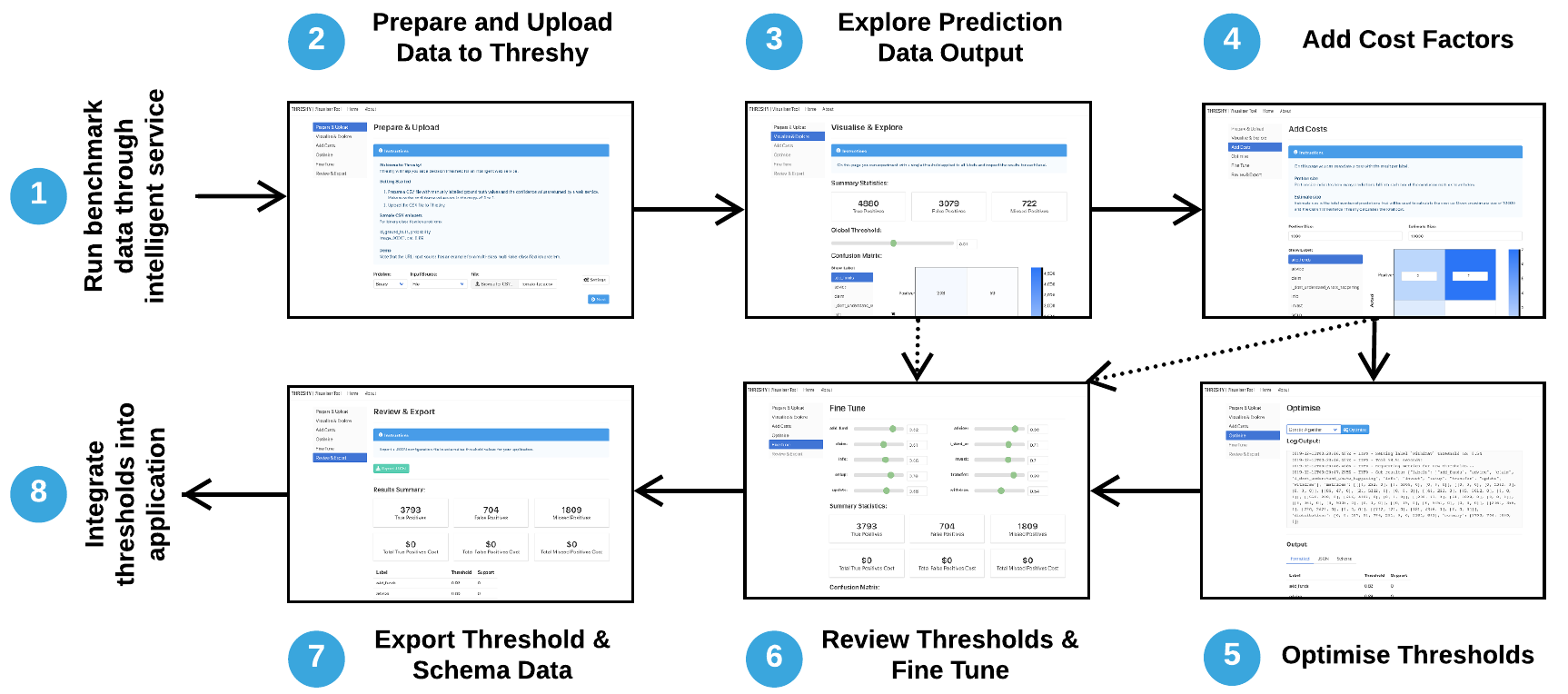}
    \caption{UI workflow for interacting with Threshy to optimise the thresholds for classification problem.}
    \label{fig:workflow}
\end{figure*}

Nina combined the predictions, their respective confidence values, and Lucy's labelled ground truths into a CSV file which was then uploaded to Threshy. Nina asked Lucy, the farmer, to assist in setting relevant costs (from a business perspective) for correct predictions and false predictions. Threshy then recommended a choice of decision threshold which Nina then fine tuned while considering the performance and cost implications.

\section{Threshy}
\label{sec:threshy}

Threshy is a tool to assist software engineers with setting decision thresholds when integrating machine-learnt components in a system in collaboration with subject matter experts. Our tool also serves as a method to inform and educate engineers about the nuances to consider when using prepackaged ML services. Key novel features are:

\begin{itemize}
    \item Automating threshold selection using an optimisation algorithm (NSGA-II \citep{996017}), optimising the results for each label.
    \item Support for user defined, domain-specific weights when optimising thresholds, such as financial costs and impact to society. This allows decision thresholds to be set within a business context as they differ between applications \citep{Drummond2006}.
    \item Handles nuances of classification problems such as dealing with multi-objective optimisation, and metric selection---reducing errors of omission.
    \item Support key classification problems including binary (e.g. email is spam or ham), multi-class (e.g. predict the colour of a car), and multi-label (e.g. assign multiple topics to a document). Existing tools ignore multi-label classification.
\end{itemize}

Setting thresholds in Threshy is an eight step process as outlined in \cref{fig:workflow}. Software engineers \circled{1}~run a benchmark dataset through the machine-learnt component to create a data file (CSV format) with true labels and predicted labels along with the predicted confidence values. The data file is then \circled{2}~uploaded for initial exploration where engineers can \circled{3}~experiment with modifying a single global threshold for the dataset. Developers may choose to exit at this point (as indicated by dotted arrows in \cref{fig:workflow}). Optionally, the engineer \circled{4}~defines costs for missed predictions followed by selecting optimisation settings. The optional optimisation step of Threshy \circled{5}~considers the performance and costs when deriving the thresholds. Finally, the engineer can \circled{6}~review and fine tune the calculated thresholds, associated costs, and \circled{7}~download generated threshold meta-data to be \circled{8}~integrated into their application.

\begin{figure}[t!]
    \centering
    \includegraphics[width=.8\linewidth]{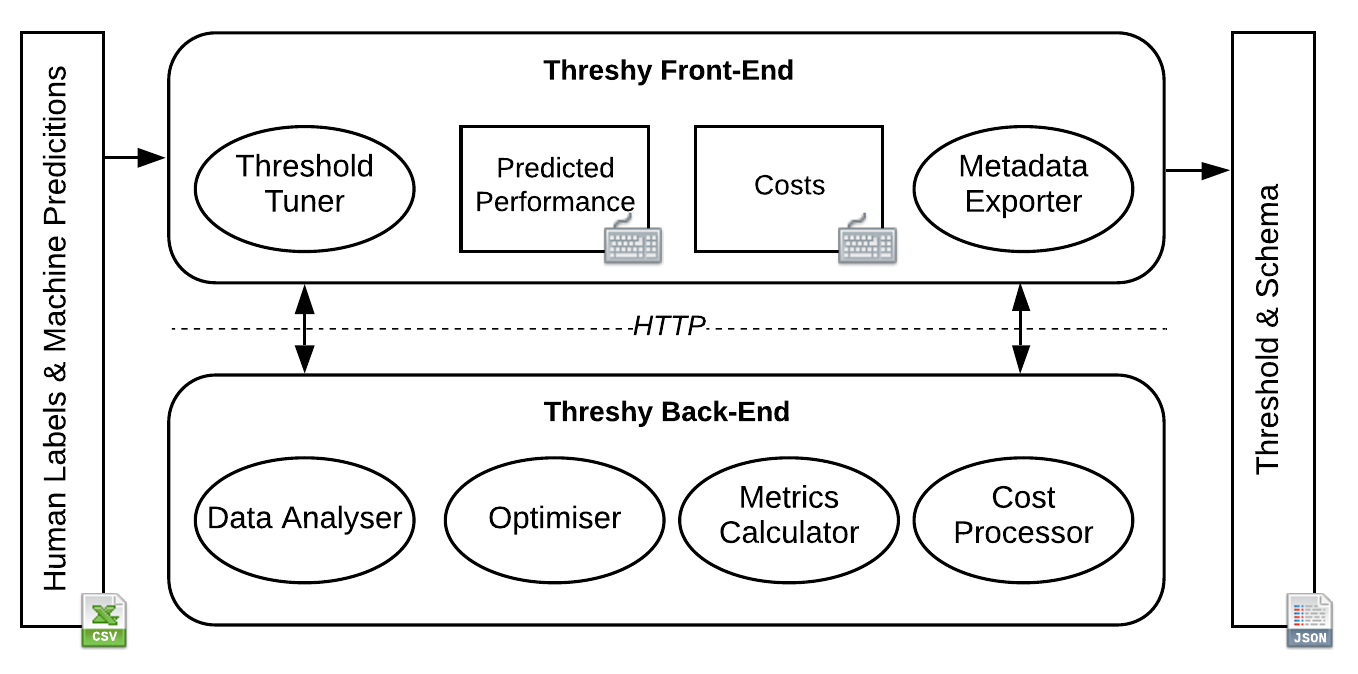}
    \caption{Architecture of Threshy.}
    \label{fig:implementation}
\end{figure}

Threshy runs a client/server architecture with a thin-client (see \cref{fig:implementation}). The web-based application consists of an interactive front-end where developers upload benchmark results---consisting of both human annotated labels and machine predictions from the intelligent service---and use threshold tuners (via sliders) to present a data summary of the uploaded information. Predicted model performances and costs are entered manually into the web interface by the developer. The Threshy back-end runs a data analyser, cost processor and metrics calculator when relevant changes are made to the front-end's tuning sliders.

The data analyser provides a comprehensive overview of confusion matrices compatible for multi-label multi-class classification problems.
When representing the confusion matrix, it is trivial to represent instances where multi-label multi-classification is not considered.
However, a more challenging case to visualise arises when you have $n$ labels and $m$ classes as the true/false matches become too excessive to visualise; $n * m * 4$ fields need to be presented.  We resolve this challenge by summarising the statistics down to three constructs: (i) number of true positives, (ii) false positives, and (iii) missed positives. This allows us to optimise against the true positives and minimise the other two constructs. Threshy is a fully self-contained repository of the tool implementation, scripting and exploratory notebooks, which is available at \url{https://github.com/a2i2/threshy}.

\section{Related work}
\label{sec:related-work}

Optimal machine-learnt decision boundaries depend on identifying the operating conditions of the problem domain. A systematic study by \citet{Drummond2006} classifies four operating conditions to determine a decision threshold: (i) the operating condition is known and the model trained matches perfectly; (ii) where the operating conditions are known but change with time, and thus the model must be adaptable to such changes; (iii) where there is uncertainty in the knowledge of the operating conditions certain changes in the operating condition are more likely than others; and (iv) where there is no knowledge of the operating conditions and the conditions may change from the model in any possible way. Various approaches to determine appropriate thresholds exist for all four of these cases, such as cost-sensitive learning, ROC analysis, and Brier scores. However, an \textit{automated} attempt to calibrate decision threshold boundaries is not considered, and is largely pitched at a non-software engineering audience. A recent study touches on this in model management for large-scale adversarial instances in Google's advertising system \citep{sculley2011detecting}, however this is only a single component within the entire architecture, and is not a tool that is useful for developers in varying contexts. Threshy provides a `plug-and-play' style calibration method where any context/domain can have thresholds automatically calibrated \textit{and} optimised for engineers. Threshy’s architecture supports a headless mode for use in monitoring workflows.

Support tools for ML frameworks generally fall into two categories. The first attempts to illuminate the `black box' by offering ways in which developers can better understand the internals of the model to improve its performance. For extensive analyses and surveys into this area, see \citep{Hohman2018VisualAI,Patel:2008:ISM:1357054.1357160}. However, a recent emphasis to probe only inputs and outputs of a model has been explored, exploring off-the-shelf models without  knowledge of its unknowns (see \cref{fig:example}) to reflect the nature of real-world development. Google's \textit{What-If Tool} \citep{DBLP:journals/corr/abs-1907-04135} for Tensorflow provides a means for data scientists to visualise, measure and assess model performance and fairness with various hypothetical scenarios and data features; similarly, Microsoft's \textit{Gamut} tool  \citep{hohman2019gamut} provides an interface to test hypotheticals on Generalized Additive Models, and a  \textit{ModelTracker} tool \citep{amershi2015modeltracker} collates summary statistics on sample data to enable visualisation of model behaviour and access to key performance metrics.

However, these tools are focused toward pre-development model evaluation and not designed for software engineering workflows. Nor are they context-aware to the overall software system they are meant to target. They are also aimed at data scientists and model builders and do not consider consistent tooling that works across development, test, and production environments. \todo[They also do not provide synthesised output for using intelligent web services with predetermined thresholds. ]{C14}. Further, certain tools are tied to specific ML frameworks (e.g., What-If and Tensorflow). Our work, instead, attempts to bridge these gaps through a context-aware, structured workflow with an automated tool targeted to software developers; our tool is designed for software engineers to calibrate thresholds and is used for intelligent service APIs in particular.%

\section{Conclusions \& Future Work}
\label{sec:conclusion}

Primary contributions of this work include Threshy, a tool for automating threshold selection, and the overall meta-workflow proposed in Threshy that developers can use as a point of reference for calibrating thresholds. \todo[
Threshy only deals with classification problems and adapting our method to other problem domains is left as future work. ]{C1: Limitations to Threshy; C3: Use of NSGA-II; C14: Synthesis} \todo[Furthermore, we plan
to evaluate Threshy with practitioners for user-acceptance and add support for code synthesis for calibrating the API responses.]{C2: More rigorous eval needed; C4: UAT}

\bibliography{fse-demo2020}
\balance

\end{document}